\documentclass{aastex61}

\usepackage{subfigure}
\usepackage{graphicx}
\usepackage{ragged2e}
\usepackage{footnote}
\usepackage{booktabs}
\usepackage{color}  
\usepackage{amsmath}
\usepackage{amssymb}
\usepackage{multirow}

\newcommand\mml[1]{{\color{black}#1}}
\begin{document}
\title{Detecting the oscillation and propagation of the nascent dynamic solar wind structure at 2.6 solar radii using VLBI radio telescopes}
\author{Maoli Ma}
\affiliation{Shanghai Astronomical Observatory, 80 Nandan Road, Shanghai,China}
\affiliation{Shanghai Astronomical Observatory, Key Laboratory of Radio Astronomy, Chinese Academy of Sciences, 10 Yuanhua Road, Nanjing, JiangSu 210033, China}
\author{Guifr\'{e} Molera Calv\'{e}s}
\affiliation{University of Tasmania, Private Bag 37, Hobart TAS 7001, Australia}
\author{Giuseppe Cim\`{o}}
\affiliation{Joint Institute for VLBI ERIC, Oude Hoogeveensedijk 4, 7991 PD Dwingeloo, The Netherlands}
\author{Ming Xiong}
\affiliation{National Space Science Center, NO.1 Nanertiao, Zhongguancun, Haidian district, Beijing}
\author{Peijia Li}
\affiliation{Shanghai Astronomical Observatory, 80 Nandan Road, Shanghai,China}
\author{Jing Kong}
\affiliation{Beijing Aerospace Control Center, Beijing, China}
\author{Peijin Zhang}
\affiliation{Institute of Astronomy and National Astronomical Observatory, Sofia, Bulgaria}
\author{Jiansen He}
\affiliation{School of Earth and Space Sciences, Peking University, Beijing, 100871, China}
\author{Lijia Liu}
\affiliation{National Astronomical Observatories, Chinese Academy of Sciences, 20A Datun Road, Chaoyang District, Beijing, China}
\author{Pradyumna Kummamuru}
\affiliation{University of Tasmania, Private Bag 37, Hobart TAS 7001, Australia}
\author{Chuanpeng Hou}
\affiliation{School of Earth and Space Sciences, Peking University, Beijing, 100871, China}
\author{Jasper Edwards}
\affiliation{University of Tasmania, Private Bag 37, Hobart TAS 7001, Australia}
\author{Qinghui Liu}
\affiliation{Shanghai Astronomical Observatory, 80 Nandan Road, Shanghai,China}
\author{Zhong Chen}
\affiliation{Shanghai Astronomical Observatory, 80 Nandan Road, Shanghai,China}
\affiliation{National Basic Science Data Center, Building No.2, 4, Zhongguancun South 4th Street, Haidian District, Beijing 100190, China}
\author{Zhanghu Chu}
\affiliation{Shanghai Astronomical Observatory, 80 Nandan Road, Shanghai,China}
\author{De Wu}
\affiliation{Shanghai Astronomical Observatory, 80 Nandan Road, Shanghai,China}
\affiliation{University of Chinese Academy of Sciences, Beijing 100049}
\author{Xu Zhao}
\affiliation{Shanghai Astronomical Observatory, 80 Nandan Road, Shanghai,China}
\affiliation{University of Chinese Academy of Sciences, Beijing 100049}
\author{Zhichao Wang}
\affiliation{Shanghai Astronomical Observatory, 80 Nandan Road, Shanghai,China}
\affiliation{University of Chinese Academy of Sciences, Beijing 100049}
\author{Songtao Han}
\affiliation{Beijing Aerospace Control Center, Beijing, China}
\author{Quanquan Zhi}
\affiliation{Beijing Aerospace Control Center, Beijing, China}
\author{Yingkai Liu}
\affiliation{Beijing Aerospace Control Center, Beijing, China}
\author{Jonathan Quick}
\affiliation{Hartebeesthoek Radio Astronomy Observatory, PO Box 443, Krugersdorp, South Africa}
\author{Javier Gonz\'{a}lez}
\affiliation{Centro de Desarrollos Tecnol\'{o}gicos, Observatorio de Yebes, Guadalajara, Spain}
\author{Cristina Garc\'ia Mir\'o }
\affiliation{Centro de Desarrollos Tecnol\'{o}gicos, Observatorio de Yebes, Guadalajara, Spain}
\author{Mikhail Kharinov}
\affiliation{Institute
of Applied Astronomy of the Russian Academy of Sciences, Kutuzova
Embankment 10, St. Petersburg, 191187, Russia}
\author{Andrey Mikhailov}
\affiliation{Institute
of Applied Astronomy of the Russian Academy of Sciences, Kutuzova
Embankment 10, St. Petersburg, 191187, Russia}
 \author{Alexander Neidhardt}
 \affiliation{12 Technical University of Munich, Geodetic Observatory Wettzell, Sackenrieder Str. 25, D-93444 Bad K$\ddot{o}$tzting, Germany}
  \author{Tiziana Venturi}
 \affiliation{Istituto di Radioastronomia Via P. Gobetti 101, 40129 Bologna,Italy}
  \author{Marco Morsiani}
 \affiliation{Istituto di Radioastronomia Via P. Gobetti 101, 40129 Bologna,Italy}
  \author{Giuseppe Maccaferri}
 \affiliation{Istituto di Radioastronomia Via P. Gobetti 101, 40129 Bologna,Italy}
\author{Bo Xia}
\affiliation{Shanghai Astronomical Observatory, 80 Nandan Road, Shanghai,China}
\author{Hua Zhang}
\affiliation{Xinjiang Astronomical Observatory, CAS 150 Science 1-Street, Urumqi, China}
\author{Longfei Hao}
\affiliation{Kunming Astronomical Observatory, CAS 150 Science 1-Street, Yunnan, China}

\correspondingauthor{Maoli Ma, mamaoli@shao.ac.cn; Guifr\'{e} Molera Calv\'{e}s, guifre.moleracalves@utas.edu.au}

\begin{abstract}
Probing the solar corona is crucial to study the coronal heating and solar wind acceleration. However, the transient and inhomogeneous solar wind flows carry large-amplitude inherent $\rm{Alfv\acute{e}n}$ waves and turbulence, which make detection more difficult. We report the oscillation and propagation of the solar wind at 2.6 solar radii (Rs) by observation of China's Tianwen and ESA's Mars Express with radio telescopes. The observations were carried out on Oct.9 2021, when one coronal mass ejection (CME) passed across the ray paths of the telescope beams. We obtain the frequency fluctuations (FF) of the spacecraft signals from each individual telescope. Firstly, we visually identify the drift of the frequency spikes at a high spatial resolution of thousands of kilometers along the projected baselines.
 They are used as traces to estimate the solar wind velocity. Then we perform the cross-correlation analysis on the time series of FF from different telescopes.
The velocity variations of solar wind structure along radial and tangential directions during the CME passage are obtained. The oscillation of tangential velocity confirms the detection of
streamer wave. Moreover, at the tail of the CME, we detect the propagation of an accelerating fast field-aligned density structure indicating the presence of magnetohydrodynamic waves. This study confirm that
the ground station-pairs are able to form particular spatial projection baselines with high resolution and sensitivity to
study the detailed propagation of the nascent dynamic solar wind structure.\\
\end{abstract}

\section{Introduction} \label{sec:intro}
\noindent
The process on how the corona is heated to be many hundreds of
times hotter than the photosphere and accelerated to form a supersonic stellar wind
is still not clear. The observation of the nascent solar wind inside 6 solar radii (Rs) is very valuable to study the acceleration model~\citep{Grall1996,Adhikari2020}. However, the near-Sun solar wind
is much more variable and structured, accompanied by the coronal mass ejection (CME), the polar plume, etc.
 In order to monitor the large scale solar wind from 1.5 to 32 Rs,
the Solar and Heliospheric Observatory (SOHO) and the on board Large Angle and Spectrometric Coronagraph (LASCO) instruments were launched
in 1995 to the Lagrange L1 to study the solar corona~\citep{Brueckner1995,Domingo1995}.
To further study the electromagnetic fields and wave-particle interactions, NASA's Parker Solar Probe (PSP) was launched in 2018 to
cross the $\rm{Alfv\acute{e}n}$ critical surface and will orbit at a perihelion of 9 Rs from the solar surface beginning in 2024~\citep{Fox2016}.
These observations have improved our understanding about the mechanisms of solar wind acceleration~\citep{Bale2019,He2021}. However, PSP will never probe as close to the Sun as the remote-sensing methods.\\
The human made spacecraft specially designed for deep space exploration provides an excellent radio source for the radio remote-sensing of the solar corona. The spacecraft operates at a high frequency band (GHz) with strong signal-to-noise ratio, which enables the observation of solar wind very close to the Sun during conjunction. Therefore, almost all the interplanetary spacecraft are used to deduce the large-scale
coronal structure. See, \cite{Patzold2016,Patzold2012} measured the total electron content using Mars Express (MEX), Venus Express (VEX) and Rosetta.
 \cite{Imamura2014} derived the radial profile of solar wind outflow speeds between 1.5-20.5 Rs from the amplitude fluctuations of the radio signal scintillation in the Akatsuki 2011 observations. \cite{Molera2017} charactered the coronal mass ejections from the MEX observations. \cite{Efimov2018} inferred the radial acceleration of slow solar wind at low heliolatitudes from two-station frequency fluctuations (FF) measurements of the Galileo spacecraft. \cite{Wexler2019} measured the speed of solar wind by examining the power spectral density of FF. \cite{Wexler2020} presented a two-component model for interpretation of the FF from Akatsuki spacecraft and determined the radial profile of slow wind speed in the extended corona using mass-flux continuity. \cite{Ma2021} measured
the radial solar wind velosity within 10 Rs with VLBI radio telescopes.\\
Tianwen-1 (TIW) is the first Chinese spacecraft exploring Mars, entering orbit around Mars on Feb.10, 2021~\citep{Zhang2022}. Meanwhile, the Mars Express spacecraft is operating in Mars orbit since early 2004~\citep{Schmidt2003,Patzold2016}.
We conducted the solar conjunction observations of TIW or MEX in 2021 with the radio telescopes from
the European VLBI Network and from the University of Tasmania. In this paper, we use the recording system of VLBI network. We do not conduct multiplying interferometer analysis, which is a standard approach for VLBI. Instead, we obtain the FF from each individual telescope, then carry the cross-correlation analysis of the FF from different telescopes to estimate the solar wind velocity.
We present the observations and data process in Sec. \ref{sec:obs and process}. Sec. \ref{sec:results} presents the propagation time and velocity measurements and Sec. \ref{sec:Conclusions} is the conclusions. \\

\section{Observations and the data process} \label{sec:obs and process}
\noindent
\subsection{Observations} \label{sec:obs}
The observations were conducted on Oct.9 2021, as indicated in Table \ref{tab:observation}.
The projected Mars' position in heliographic latitude and Carrington longitude are 51$^{\circ}$ and 258$^{\circ}$, with the heliocentric distance 2.6 Rs from the center of the Sun.
TIW and MEX were observed in the same beam by the EVN telescopes of Hartebeesthoek 25 m (Hh), Zelenchukskaya 32 m(Zc), Badary 32 m(Bd), Medicina 32 m (Mc), Yebes 40 m (Ys), and Yarragadee 12 m (Yg) antenna of the University of Tasmania.
The observation of TIW continued from 06:50 to 13:00, and MEX ended at 08:30. Due to the limitation of visibility, Mc and Ys participated the observations later at 07:40, and Bd ended earlier at 09:30.
The observation covered the eruption, passage typical of a CME. The CME was a halo in SOHO LASCO C2 and C3. The associated eruption followed the M1.6 class flare from AR 2882 and was characterized by significant dimming, an EUV wave and post-eruptive arcades seen mostly to the West from AR 2882 in SDO AIA 193, 304, 171 and in EUVI A 195 starting at 2021-10-09T06:33Z\footnote{https://kauai.ccmc.gsfc.nasa.govs/DONKI/view/CMEAnalysis/17926/3}.\\
The radio telescopes observed the X-band (8.4 GHz) downlink signals from the spacecraft.
TIW was working in safe mode with low gain antenna on board and transponder non-coherent mode. The equivalent isotropically radiated power of the transmitter on board is only $\sim$26 dBW. The Allan variance of the onboard oscillator is $10^{-12}$ per second, allowing us to identify the FF caused by the interplanetary scintillation from the equipment noise. MEX was operating in a closed-phase locked loop with one of the antennas of the European Space Agency's tracking stations.\\

\begin{table}[htb]
\startlongtable
\begin{deluxetable}{c c c c c c c}
\tablecaption{Specifications of the observations on Oct.9 2021: time, the projected Mars¡¯ position, stations and targets.\label{tab:observation}}
\tablehead{
\colhead{Time} & \colhead{heliocentric distance [$\rm R_{s}$]}& \colhead{heliographic latitude [$^{\circ}$]}&Carrington longitude [$^{\circ}$]&\colhead{Stations}&\colhead{Targets}}
\startdata
  UTC 06:50-13:00   &    2.62   &  51 & 258 &     Hh,Zc,Bd,Mc,Ys  &     TIW\\
  UTC 06:50-08:30   &    2.62   &  51 & 258 &     Hh,Zc,Bd,Mc,Ys,Yg  &  MEX \\
\enddata
\end{deluxetable}
\end{table}

\subsection{Measure the frequency fluctuations} \label{sec:pro}
\begin{figure}[htb]
\centering
\includegraphics[width=1.0\textwidth]{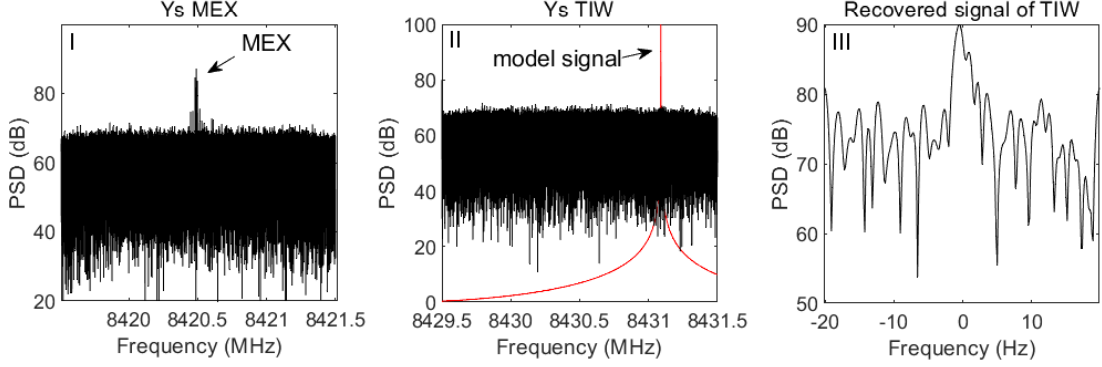}
\caption{PSD of MEX and TIW at Ys station at 08:30 on Oct.9. I. MEX. II. TIW and the dynamic signal model. The main carrier of TIW is obscured by the noise. III. recover the signal of TIW with local correlation.\label{fig:Yssignal}}
\end{figure}

\begin{figure}[htb]
\centering
\subfigure[TIW]{\includegraphics[height=10cm,width=20cm]{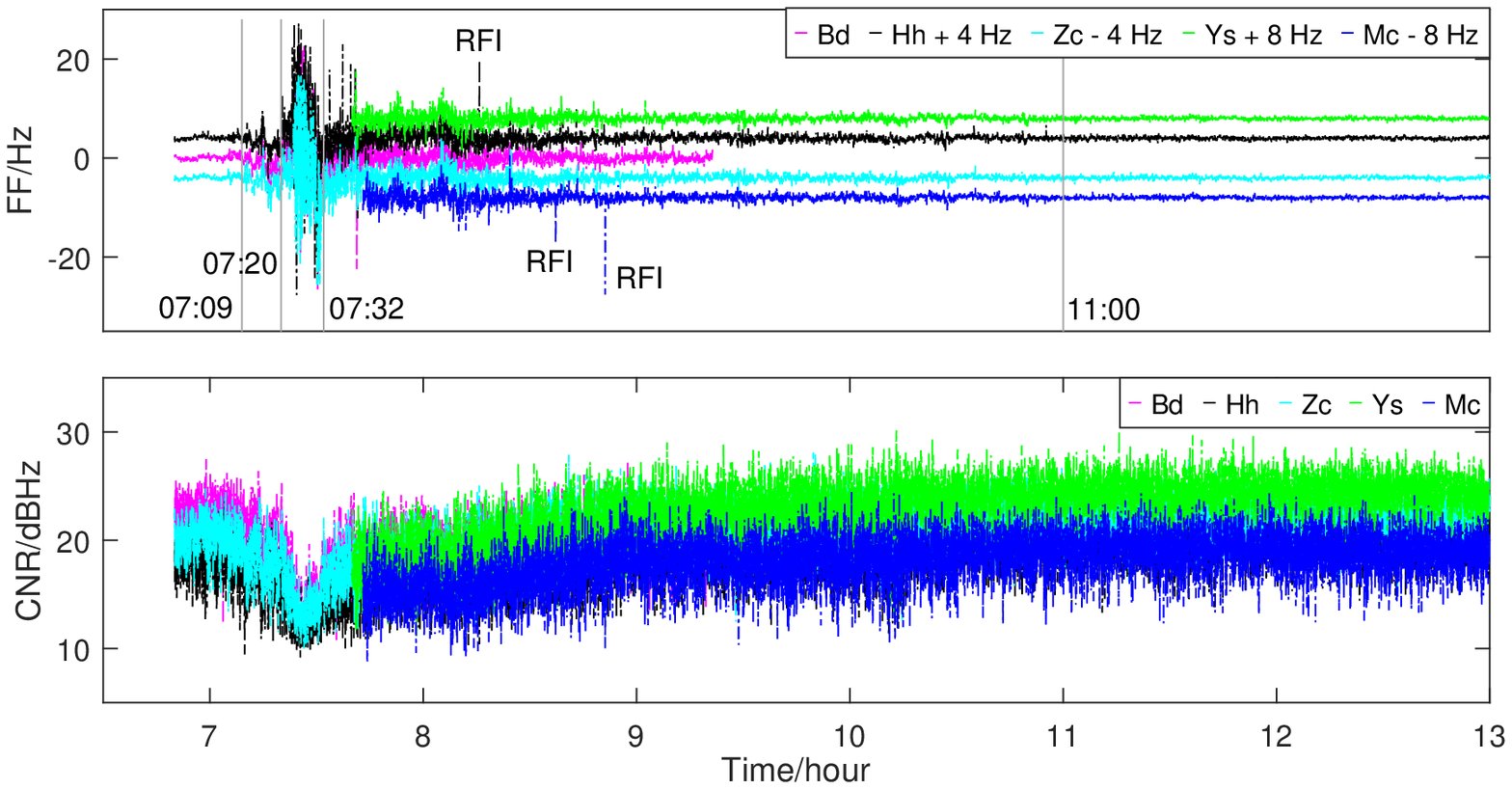}}
\subfigure[MEX]{\includegraphics[height=10cm,width=20cm]{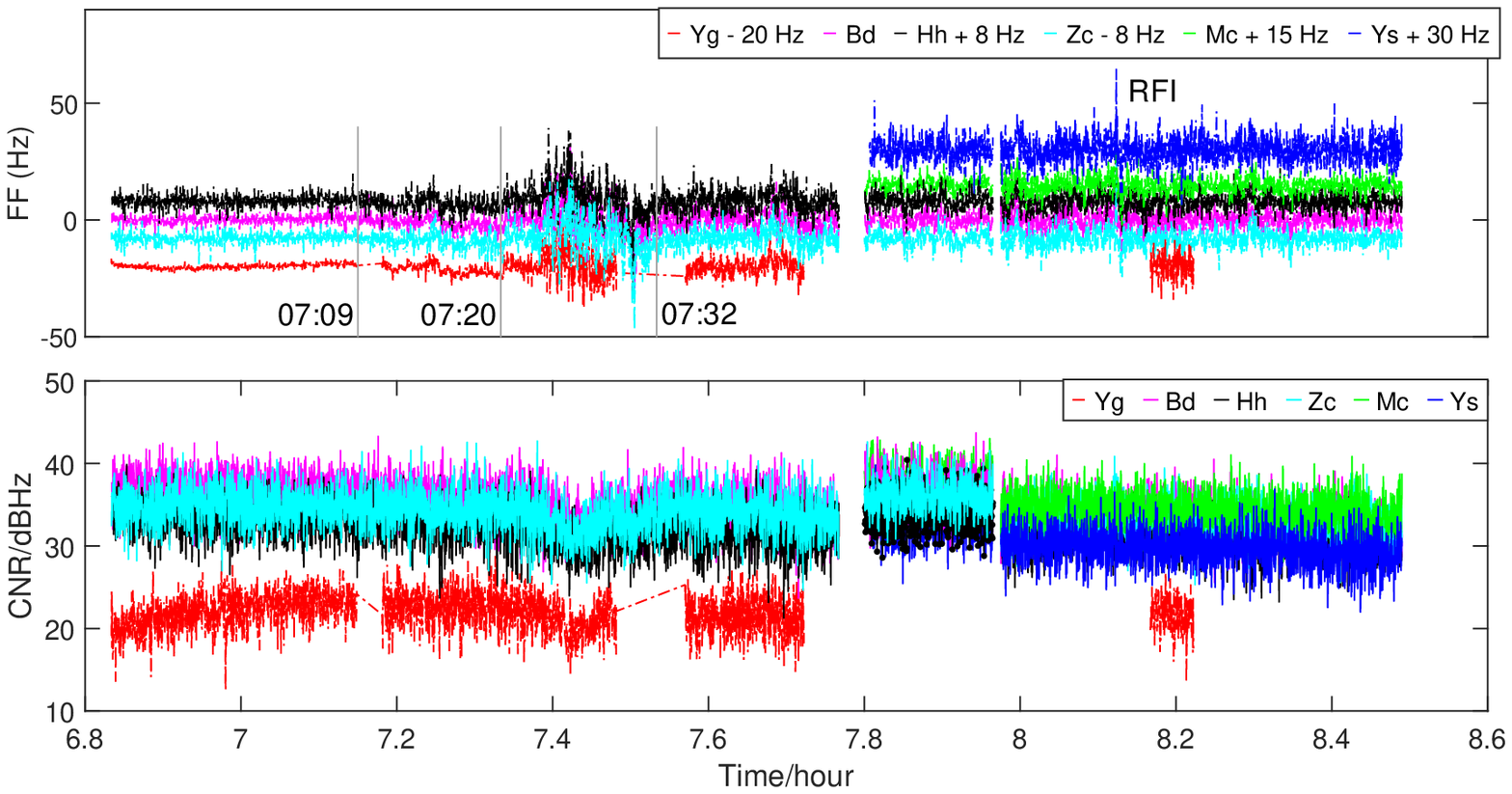}}
\caption{The FF and CNR of TIW (a) and MEX (b) \mml{from individual telescopes}. The FF is shifted by a constant value given in the legend.
\mml{The CME entered the LASCO-C2 field of view at 07:09} \mml{The CME
front arrived the projected Mars's position around 07:20. The effects of CME on the signal
weaken after 07:32 and fade away after 11:00. In (a), the frequency jumps appear at 08:37:16 and 08:51:12 only at Mc are the RFIs. In (b), a RFI appear at the same time
08:07:23 from Hh, Zc, Bd, Ys and Mc stations when observing MEX.}}
\label{fig:TIWandMEX}
\end{figure}

We have applied three methods to obtain the time series of the FF of the downlink spacecraft signal.
 Fig.\,\ref{fig:Yssignal} is the power density spectra (PSD) of signals from MEX and TIW at Ys station at 08:30 on Oct.9. The carrier to noise ratio (CNR) is defined as the ratio of the main carrier power to the noise density.
 The CNR of MEX is about 30 dBHz.
 Due to the low gain antenna used by TIW, the received signal is extremely weak, less than 10 dBHz, totally obscured by the noise.
We use the local correlation method to extract the frequency of the weak signal~\citep{Ma2021a}.
 To mitigate the CNR loss due to frequency smearing caused by Doppler shift, the local correlation method compensates for the Doppler shift of the main carrier with a signal propagation
model constructed by the kinematics of the spacecraft and the onboard transmitted frequency. Only the signal that are dynamically similar to the model can be recovered as a detectable one as Fig.\,\ref{fig:Yssignal} III.\\
The CNR of the signal after the local correlation is about 20 dBHz, enables the frequency measurements.
Since the signal of TIW could not be resolved in the PSD of the raw data, we utilize its CNR after the local correlation in the following pictures for analysis. For the stronger signal of MEX, the CNR in the PSD of the raw data is applied.\\
Owing to lacking the transmission frequency information of MEX, we use two other methods instead to process them. The data recorded at Yg station is processed
with the high spectral resolution multi-tone spacecraft Doppler tracking
software developed by \cite{MoleraCalves2021}. And we use the instantaneous Doppler frequency method to obtain the received frequency of MEX observed at other EVN telescopes~\citep{Zheng2013}.
A polynomial fit is applied to the frequency time series to determine the variation tendency, then subtracted from the frequency time series to generate an FF time series about zero. For the residual frequency measured from local correlation, a 2-order fit is used~\citep{Ma2021a}. For the received frequency of MEX, a 6-order fit is used to remove away the Doppler shift. Due to the weak signal of TIW, the integration time of the frequency fluctuations is set to 2 s. Fig.\,\ref{fig:TIWandMEX} (a) and (b) present the frequency fluctuations and the CNR for TIW and MEX on Otc. 9, respectively. The observations mode for Yg antenna includes observing the target for 19 minutes with a 1-minute break for the repointing and calibration. Some longer data gaps were caused for a switch in operations of the spacecraft between one to two-way mode. At Yg antenna we only process data in two-way mode. For other EVN telescopes, we carefully delete the data around the switch time. A detailed analysis on the FF is presented in Sec. \ref{subsec:ff}.\\

\section{Results and analysis} \label{sec:results}
\subsection{The effect of the density inhomogeneities on spacecraft signals}\label{subsec:ff}
\begin{figure}[htb]
\centering
\includegraphics[width=1.0\textwidth]{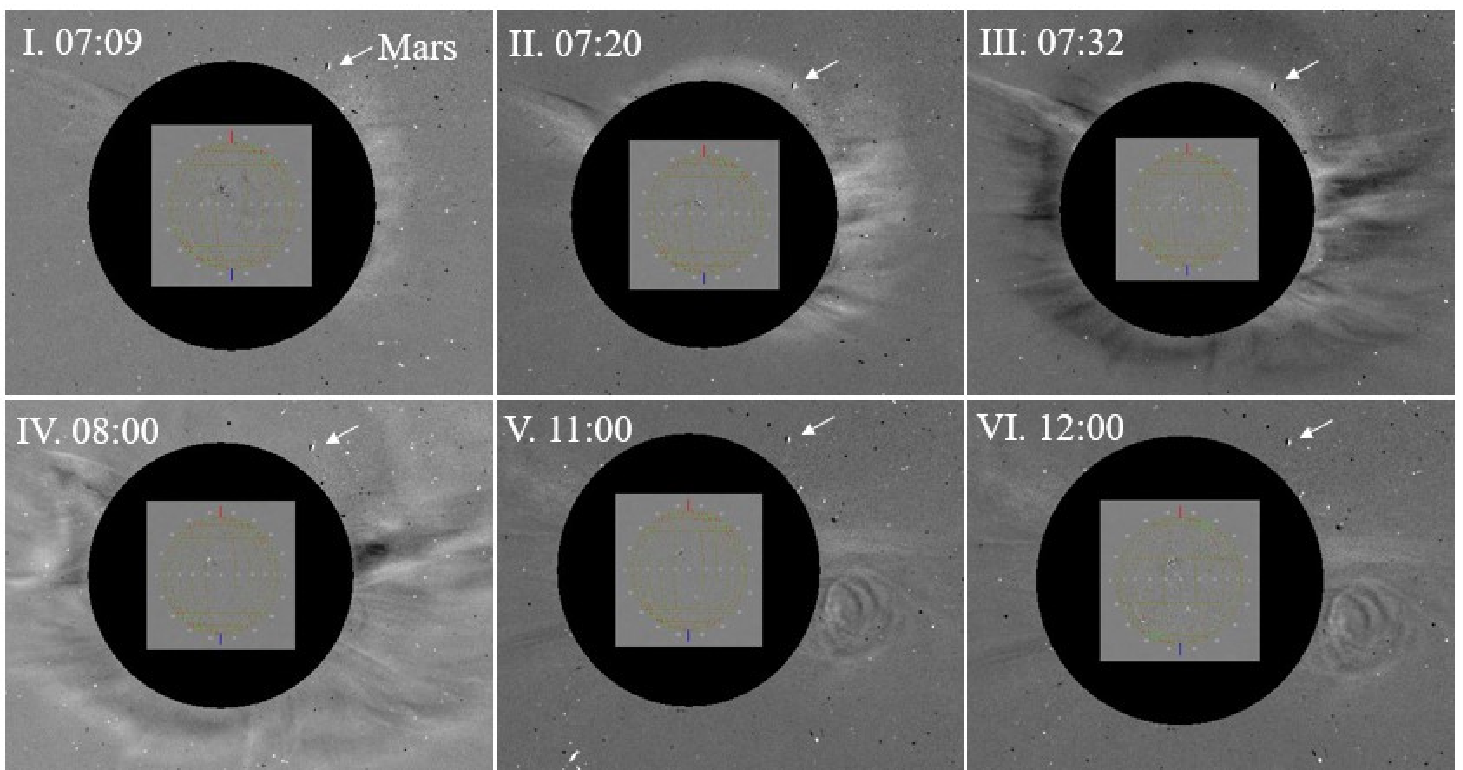}
\caption{The LASCO pictures. I. The CME entered the LASCO-C2 field of view at 07:09. II,III. The CME front arrived at the projected Mars' position. IV. The CME weaken after 08:00. V,VI. The fade away of CME.\\
 \label{fig:LASCOpic}}
\end{figure}

The CME entered the LASCO-C2 field of view at 07:09 (Fig.\,\ref{fig:LASCOpic} I). The CME
front arrived at the projected Mars' position around 07:20 and lasted until 07:32 (Fig.\,\ref{fig:LASCOpic} II and III).
It spreaded out (Fig.\,\ref{fig:LASCOpic} IV), and almost faded away after 11:00.
The differences in images between 11:00 and 12:00 were not distinguishable (Fig.\,\ref{fig:LASCOpic} V and VI).
In Fig.\,\ref{fig:TIWandMEX}, we can see the effects of CME on both TIW and MEX signals. The FF of the background solar wind are -1$\sim$1 Hz.
From 07:06, stronger FF distinctly appear above the background field, and become very severe
between 07:20-07:32, with the fluctuations up to -30$\sim$30 Hz and the CNR decreasing about 5$\sim$8 dBHz. The effects on signal
weaken after 08:00 and fade away after 11:00.
Here we call the stronger FF frequency 'spikes'
which are distinguished because of the density
contrast between the transient inhomogeneities and the ambient flow.\\
The onboard or ground systems could result in some abnormal radio frequency interference (RFI) as well.
The simultaneous observations of multiple stations and spacecraft enable us to distinguish the spikes caused by solar plasma from the RFI.
Sometimes, the RFIs appear only on one station but not on another. See, the frequency jumps appear at 08:37:16 and 08:51:12 only at Mc. We also find a RFI appear at the same time
08:07:23 at Hh, Zc, Bd, Ys and Mc stations observing MEX, however, we don't see the corresponding RFI from TIW. This RFI is caused likely by some factor on MEX. Those are excluded in the following analysis.\\
\subsection{Measure the propagation time and velocity of the solar wind from visual spikes}\label{subsec:inho}
\begin{figure}[htb]
\centering
\includegraphics[width=1.0\textwidth]{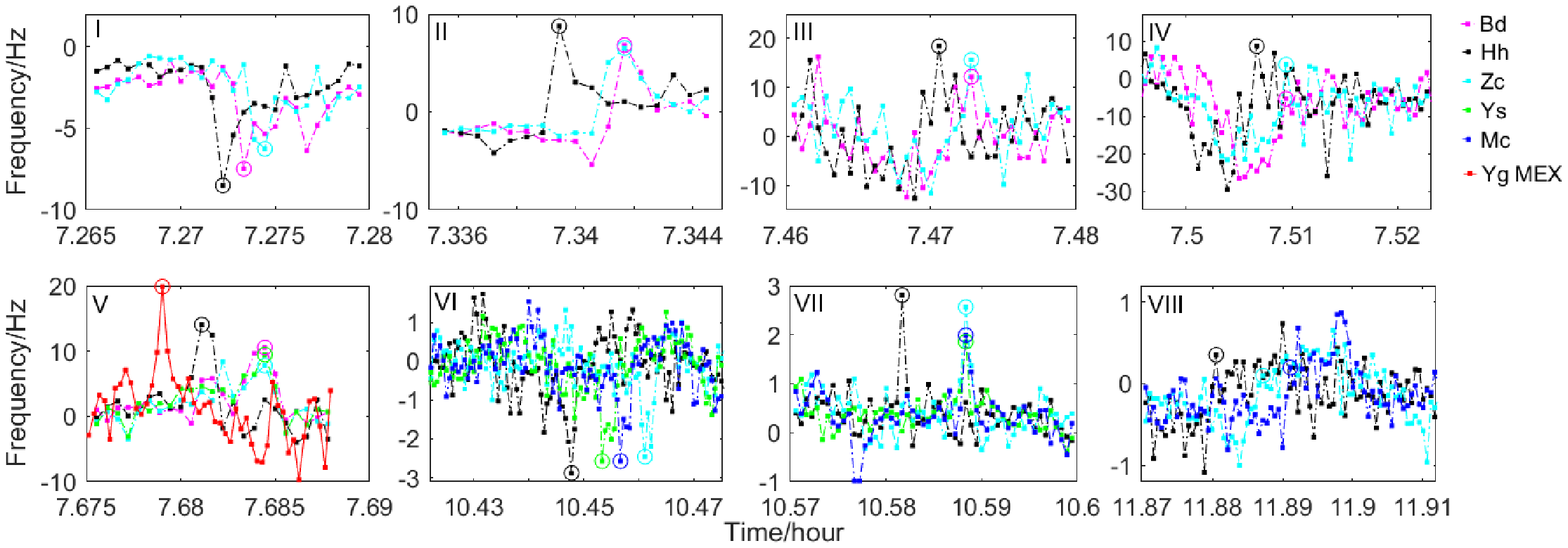}
\caption{The spikes in frequency fluctuations at different moments \mml{from individual telescopes}. I,II,III,IV. Spikes among Hh, Bd and Zc from TIW. V. Spikes among Yg from MEX, Hh, Bd, Zc from TIW. VI,VII. Spikes among Hh, Zc, Ys and Mc from TIW. VIII. Spikes from Hh, Zc and Mc from TIW.\\
 \label{fig:LagTIWMEX}}
\end{figure}

By inspection of the spikes at different stations, the appearance time of spikes is different.
The absolute values of the spikes are larger than the ambient FF. We pick up the typical points (such as the peaks of the spikes) by visual comparison and mark with '$\circ$' to calculate the time lag.
In Fig.\,\ref{fig:LagTIWMEX} I, a spike appears at Hh at 07:16:20 with the jumping value -6.8 Hz. The recording time for the similar spikes at Bd and Zc is 07:16:24 and 07:16:28, respectively. Their delays relative to Hh are 4 s for Bd-Hh  and 8 s for Zc-Hh.
In Fig.\,\ref{fig:LagTIWMEX} V, the spikes appear from MEX as well as TIW.  We prefer to display the spikes from Yg MEX and TIW as comparison, for they are enough to show the lags.
In Fig.\,\ref{fig:LagTIWMEX} IV, the spikes with $\sim$30 Hz appear on TIW around 07:30. We also find the similar spikes around 07:30 on MEX and not show here.\\

The different occurrence time of the spikes at different stations provides the visual evidences of the drift of the scintillation pattern among the ray paths.
By differing the different occurrence time we obtain the propagation time of the inhomogeneities.
To study the propagation velocity, we have projected the baselines between the station-pairs onto the sky-plane in the heliographic coordinates, then
calculated the components in the radial $\overrightarrow{P}_{rad}$ and latitudinal $\overrightarrow{P}_{tan}$ from the sun~(see appendix).
 The time lags, $\overrightarrow{P}_{rad}$, $\overrightarrow{P}_{tan}$, and the related velocity $\upsilon_{rad}$, $\upsilon_{tan}$ of the spikes between two stations are listed in Table \ref{tab:lag}.
 Typically, the time lags between the Hh/Ht or Yg related radial baselines are similar,
 except two moments of 10:27:12 and 11:52:50.
 At 10:27:12, we find the indispensable lags difference along the tangential direction, with 12 s, 16 s and 28 s detected by Mc-Ys, Zc-Mc and Zc-Ys. The corresponding $\upsilon_{tan}$ are 66, 89 and 79 km s$^{-1}$ polewards. At 11:52, the lag between Mc-Zc is 6 s, and $\upsilon_{tan}$ is 210 km s$^{-1}$ equatorwards. We don't calculate the lag of Ys related baseline for the spike of Ys is not distinct enough to find the typical point.\\
The lag and velocity estimation by visual comparison of spikes at different stations can give us a straightforward understanding about the propagation of the solar wind density structures. To depict the velocity variation during the whole observation, we then perform the cross-correlation analysis on the time series of FF~\citep{Ma2021}.\\
 \begin{table}[htb]
\startlongtable
\begin{deluxetable}{c c c c c c c c}
\tablecaption{The lags, projected distance and velocity of the solar wind measured on Oct.9: the spikes, the spike happen at the first station, the spike happen at the second station, the time lag of the spike between the two stations,
 the radial distance $P_{rad}$, the tangential distance $P_{tan}$, the radial velocity $v_{rad}$, the tangential velocity $v_{tan}$.\label{tab:lag}}
\tablehead{
\colhead{Spikes} & \colhead{Time} & \colhead{Time}&\colhead{lag} &\colhead{$P_{rad}$}&\colhead{$P_{tan}$}&\colhead{$v_{rad}$}&\colhead{$v_{tan}$}\\
\colhead{[Hz]} & \colhead{} & \colhead{}&\colhead{[s]} &\colhead{[km]}&\colhead{[km]}&\colhead{[km s$^{-1}$]}&\colhead{[km s$^{-1}$]}}
\startdata
-6.8  &Hh   07:16:20&  Zc 07:16:28&8 & 4235 & 1550  &529&/\\
 -6.8  &Hh   07:16:20&  Bd 07:16:24&4 & 4218 &4212  &1054&/\\
8.1  &Hh   07:20:22&  Zc 07:20:30&8 & 4230 & 1552 &528&/\\
8.1  &Hh   07:20:22&  Bd 07:20:30&8 & 4200 &4205 &525&/\\
25.2  &Hh   07:28:14&  Zc 07:28:22&8 &4220 &1555&527&/\\
25.2  &Hh   07:28:14&  Bd 07:28:22&8 &4190 &4180&523&/\\
30.8  &Hh   07:30:9.5& Zc 07:30:18.5&9 &4225 &1560   &469(MEX)$^{a}$&/\\
30.8  &Hh   07:30:9.5& Bd 07:30:18.5&9 &4178 &4179  &464(MEX)$^{a}$&/\\
30.8  &Hh   07:30:24& Zc 07:30:34&10 &4230 &1556  &423&/\\
30.8  &Hh   07:30:24& Bd 07:30:34&10 &4190 &4178   &419&/\\
11.9  &Hh   07:40:52&  Zc 07:41:04&12 &  4200&1560 &350&/\\
11.9  &Hh   07:40:52&  Ys 07:41:04&12 &  4150&-170 &345&/\\
11.9  &Hh   07:40:52&  Bd 07:41:04&12 &  4160&4155  & 346&/\\
11.9  &Yg   07:40:44&  Zc 07:41:04&20 & 4650&-1000  &232&/\\
11.9  &Yg   07:40:44&  Ys 07:41:04&20 & 4600&-2500  &230&/\\
11.9  &Yg   07:40:44&  Bd 07:41:04&20 & 4590&1500  &230&/\\
-2.5  &Hh   10:26:52&  Ys 10:27:12&20 & 4393& -900  &219&/\\
-2.5  &Hh   10:26:52&  Mc 10:27:24&32 &  4465&400  &139&/\\
-2.5  &Hh   10:26:52&  Zc 10:27:40&48 & 4160&1320   &86&/\\
-2.5  &Ys   10:27:12&  Mc 10:27:24&12 & 100& 800  &/&66\\
-2.5  &Mc   10:27:24&  Zc 10:26:40&16 & -300&1428 &/&89 \\
-2.5  &Ys   10:27:12&  Zc 10:27:40&28 & -200&2221 &/&79\\
2.5  &Hh   10:34:54&  Zc 10:35:18&24 & 4160   &1200&173&/ \\
2.5  &Hh   10:34:54&  Ys 10:35:18&24 & 4416   &-905&184&/ \\
2.5  &Hh   10:34:54&  Mc 10:35:18&24 & 4475   &-120&186 &/\\
0.5  &Hh   11:52:50&  Zc 11:53:12&22 & 4200   &1000&190&/\\
 0.5  &Hh   11:52:50&  Mc 11:53:28&30 & 4550   &-200&150&/\\
0.5  &Zc   11:53:12&  Mc 11:53:28&6 & 380   &-1258&/ &-210\\
\enddata
\end{deluxetable}
\vspace{8pt}
\footnotesize{$^a$ This measurement is from the solar conjunction observation of MEX.}\\
\end{table}

\subsection{Measure the propagation time and velocity of the solar wind from cross-correlation}\label{subsec:corss_ff}
\begin{figure}[htb]
\centering
\includegraphics[width=1\textwidth]{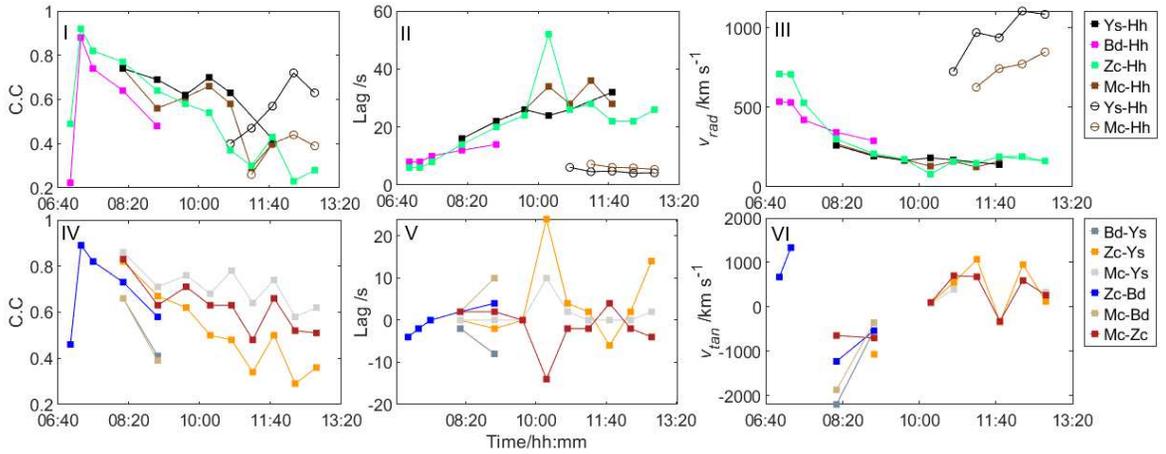}
\caption{The correlation coefficient, lag and velocity of the solar wind. The circle plots with Ys-Hh and Mc-Hh relate to the
 'fast stream'. I,II and III, the C.C, lag and velocity along the radial direction. IV,V and VI, the C.C, lag and velocity along the tangential direction.\label{fig:cclagv}}
\end{figure}

\begin{figure}[htb]
\centering
\subfigure[Hh-Ys]{\includegraphics[width=0.8\textwidth]{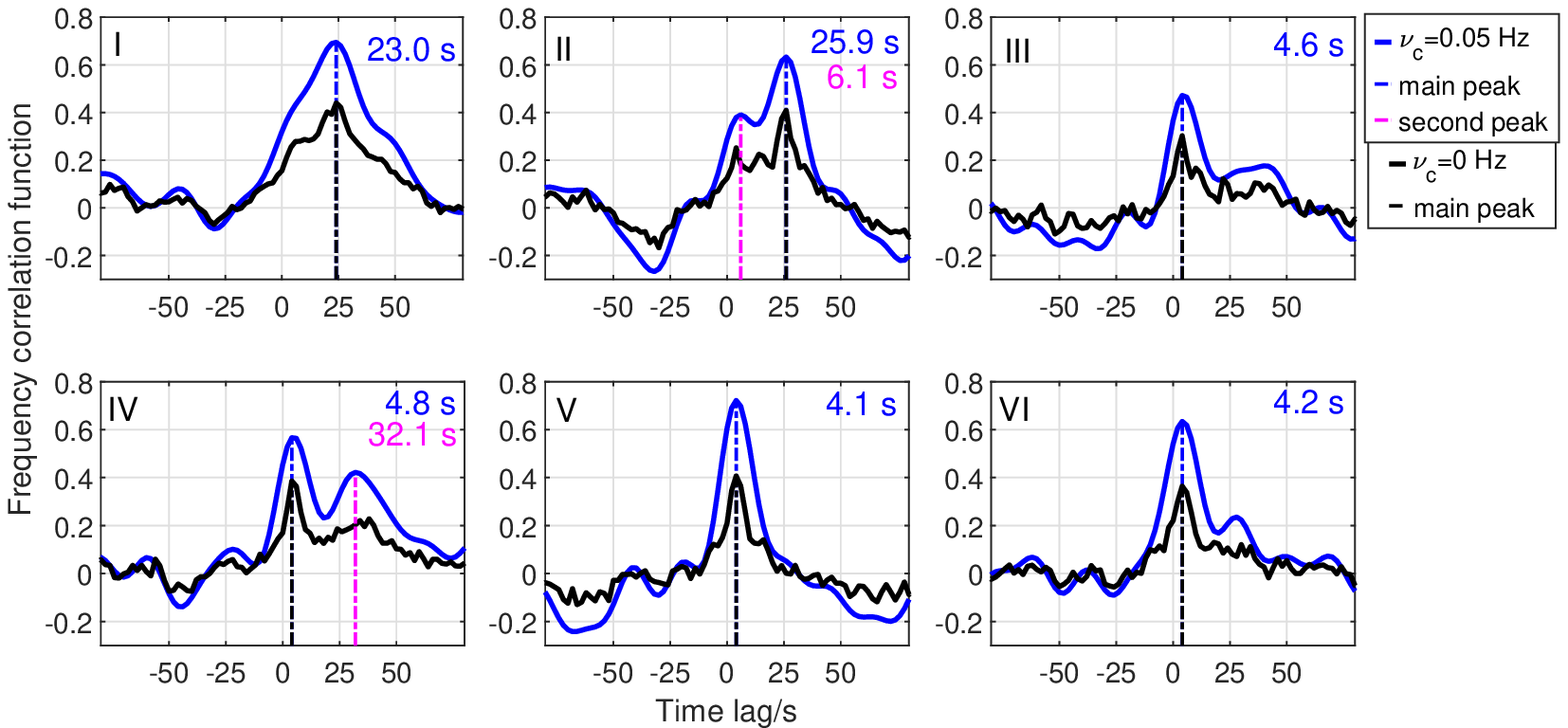}}
\subfigure[Hh-Mc]{\includegraphics[width=0.8\textwidth]{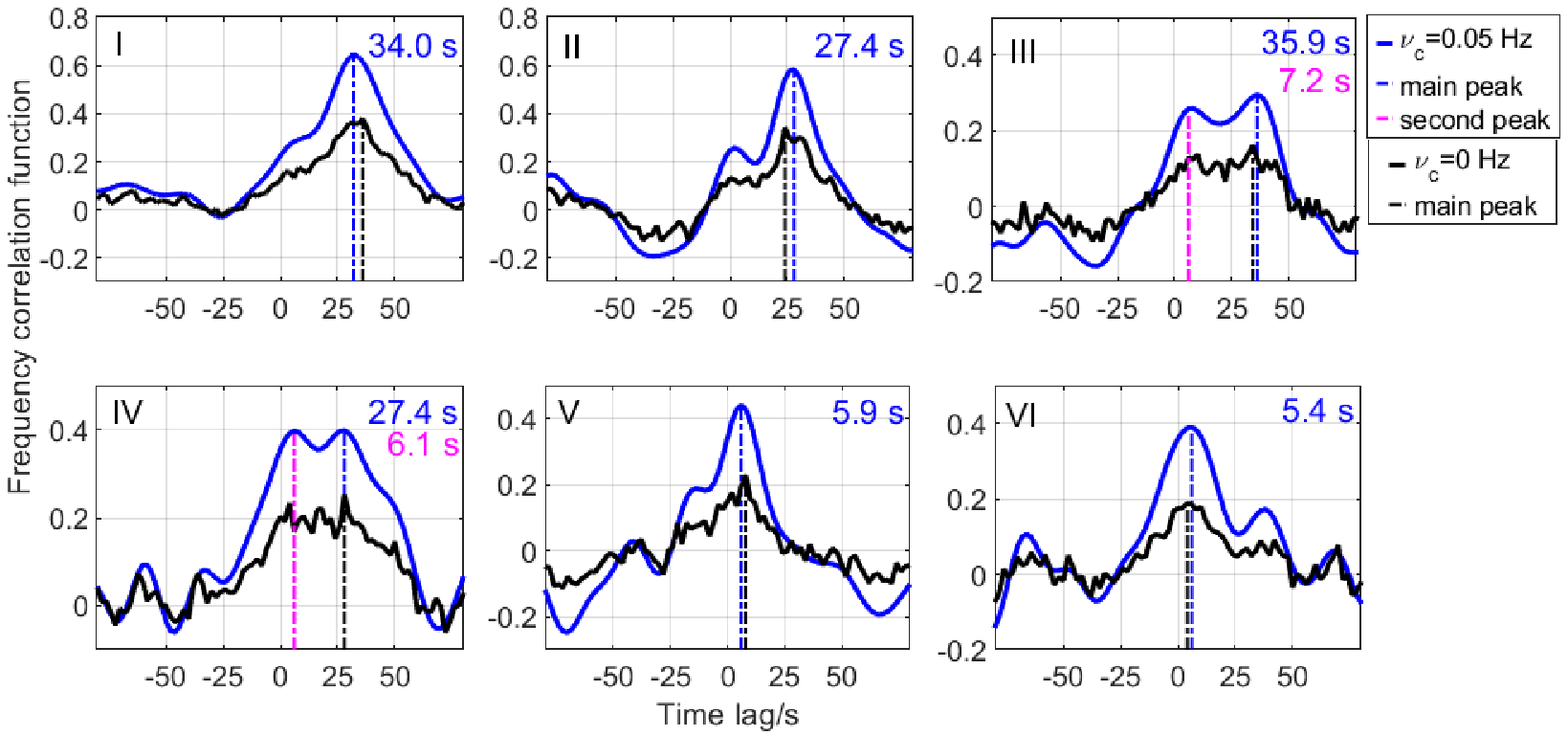}}
\subfigure[Hh-Zc]{\includegraphics[width=0.8\textwidth]{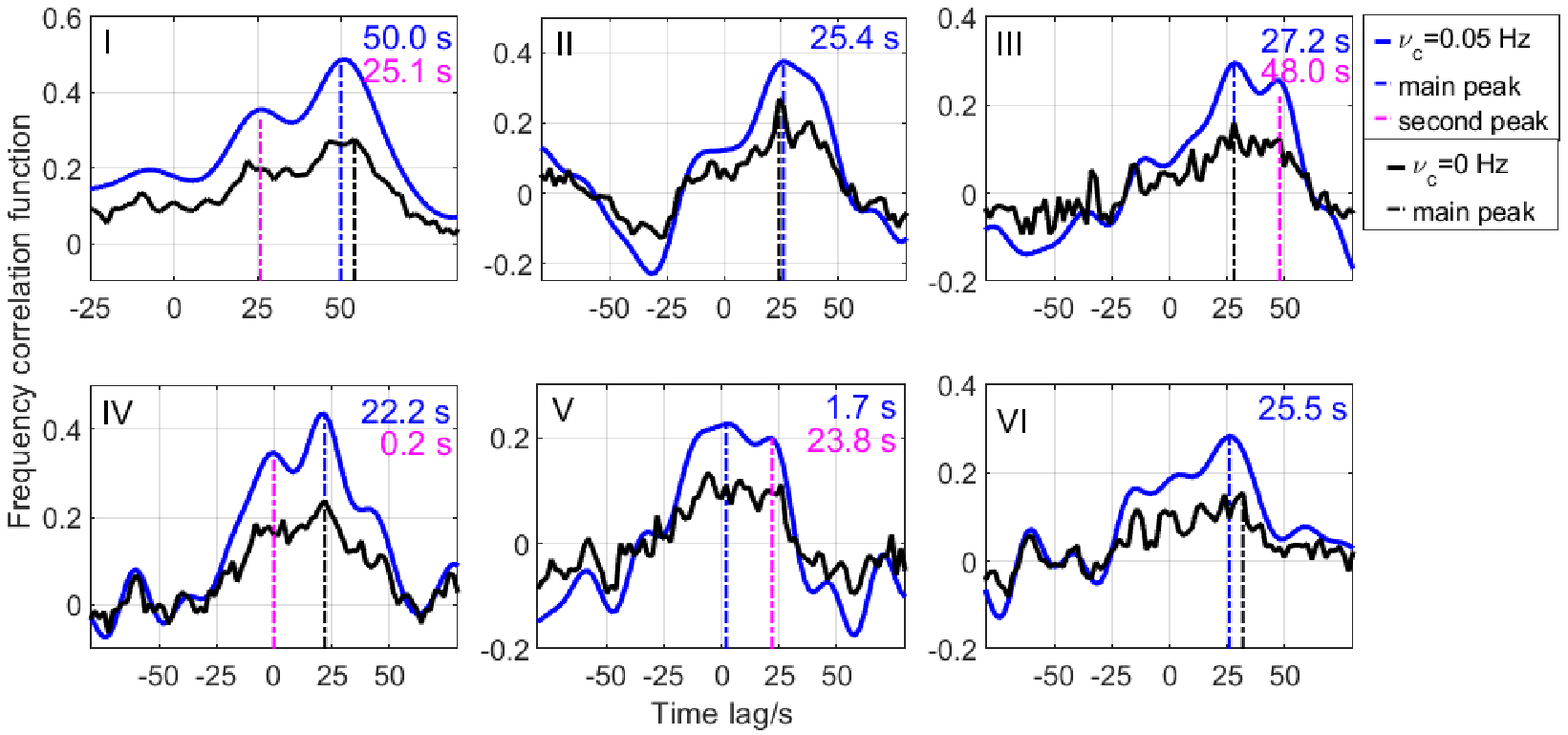}}
\caption{The cross correlation function of frequency fluctuations of Hh-Ys(a), Hh-Mc(b) and Hh-Zc(c).I.10:00-10:30. II.10:30-11:00.  III.11:00-11:30. IV.11:30-12:00. V.12:00-12:30. VI.12:30-13:00.}
\label{fig:CCfreq}
\end{figure}

We divide the FF into 12 continuous time series with a mean duration of 30 mins, thus, 06:50-07:05, 07:05-07:20,
07:20-07:44, 07:44-08:40, 08:40-09:20, 09:20-10:00, 10:00-10:30, 10:30-11:00, 11:00-11:30, 11:30-12:00, 12:00-12:30 and 12:30-13:00.
To balance the scintillation pattern and the instrument noise, the cutoff frequency $\nu_{c}$ is set to 0.05 Hz. It is the most
suitable to retain the scintillation from CME and filter the interferences from instrument noise. In order to improve the resolution of the time lag, we take a 2-order polynomial fitting on 6 points around the peak of CCFs. Then we obtain the cross-correlation coefficient (C.C) and the related lag of the peak from the fit curve. The error of lag
is obtained through analysing the uncertainties of the fit coefficients.\\
In Fig.\ref{fig:cclagv}, the C.C before the eruption of CME are below 0.5 on the baselines of Bd-Hh, Zc-Hh and Zc-Bd. They suddenly rise to 0.9 owing to the eruption of CME, then gradually decrease with the decline of the CME. The lags on the radial baselines are larger than 8 s, and $\upsilon_{rad}$ gradually decelerates from $\sim$500 to $\sim$100 km $\rm{s^{-1}}$.\\
After 10:30, accompanying the fading away of the CME, we clearly see the presence of two solar wind streams crossing the lines of sights.
Fig.\ref{fig:CCfreq} presents the CCFs of Ys-Hh, Mc-Hh and Zc-Hh between 10:30-13:00. Fig.\ref{fig:CCfreq}(a) II shows a 'bump' with two distinct peaks between 10:30-11:00 on Ys-Hh. The main peak with lag of 25.9 s relates to the 'slow stream' corresponding to the CME, and the other with lag of 6.1 s relates to a 'fast stream'.
After 11:00, the scintillation pattern then is dominated by the fast stream. The C.C of
cross-correlation peak relating to the fast stream is up to 0.6, stronger than C.C of the tail of CME. And the peak relating to the CME almost
disappears after 12:00. We also see the interaction of CME and the fast stream on Mc-Hh (Fig.\ref{fig:CCfreq}(b)). The time lag of the fast stream is between 5.4$\sim$7.2 s between 11:00-13:00, with C.C of the peak up to 0.4.
We fail to detect the fast stream on the baseline of Zc-Hh (Fig.\ref{fig:CCfreq}(c)).
Instead, with the
tangential distance of $\sim$ 1500 km at Zc-Hh, the
scintillation pattern in Fig.\ref{fig:CCfreq}(c) I includes both the radial and tangential components. The lag of the main peak is 50.0 s, matching with the lag of 48 s measured at 10:26:52 in Table \ref{tab:lag}.
The lag of the second peak is 25.1 s, consistent with the radial components measured at other time series.\\
The fast streams display an acceleration progress, 725 $\sim$ 1106 km $\rm{s^{-1}}$ obtained from Ys-Hh, 626 $\sim$ 848 km $\rm{s^{-1}}$ from Mc-Hh.
 To further verify the fast stream, we try not to use the filter on FF ($\nu_{c}$=0 Hz). The lags from $\nu_{c}$=0 Hz at Ys-Hh are consistent with $\nu_{c}$=0.05 Hz in the order of 1 s with C.C of $\sim$ 0.4. At Mc-Hh, the lags of the fast stream are between 2$\sim$8 s with C.C of $\sim$ 0.2.
It indicates the fast stream propagating better along Ys-Hh.\\
 In Fig.\ref{fig:cclagv} VI, the direction of $\upsilon_{tan}$ reverses for 4 times with the evolution of the CME. All baselines indicate the equatorwards component between 07:40-09:25. The $\upsilon_{tan}$
around 09:00 are -350, -532, -532 and -698 km $\rm{s^{-1}}$ measured from the Mc-Bd, Zc-Bd, Mc-Ys and Mc-Zc.
After 10:00, $\upsilon_{tan}$ exhibites elegant large scale sinusoidal wavelike motions.
A definitively
 poleward component appears between 10:00-10:30 with 102.5, 93, and 79 km $\rm{s^{-1}}$ measured from Mc-Zc, Zc-Ys and Mc-Ys.
 Then $\upsilon_{tan}$ turns to equatorwards between 11:30-12:00,
 and polewards again after 12:00. Two reverses of $\upsilon_{tan}$ happened between 10:00-10:30 and 11:30-12:00 match
  the measurements from visual spikes in Table \ref{tab:lag}. The deflection to the north pole of the sun results in the spikes of -2.5 Hz (redshift) at 10:27 in Fig.\,\ref{fig:LagTIWMEX} VI. Another deflection to the ecliptic plane results in the spikes of 0.9 Hz (blueshift) at 11:52 in Fig.\,\ref{fig:LagTIWMEX} VIII. These distinct drifting spikes make our measurements of the oscillation more convincing.\\


\subsection{Discussion}\label{subsec:discussion}
\begin{figure}[htb]
\centering
\includegraphics[width=1.0\textwidth]{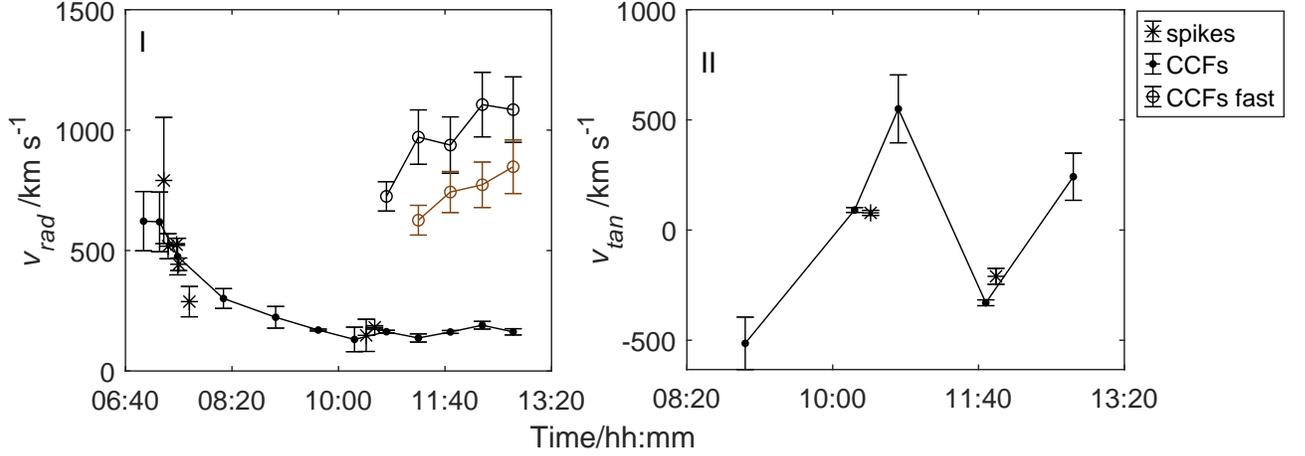}
\caption{The velocity obtained from the visual spikes and the cross-correlation. I. Radial velocity. II. Tangential velocity. \label{fig:velocity_m1a09x}}
\end{figure}

The observations from multi telescopes give us an opportunity to evaluate the consistency and rationality of the velocity. We calculate the mean velocity and the standard deviation (STD) at each moment or same time series from the different baselines. The error bars are the STD. For the velocity measured from single baseline, e.g., the field-aligned fast stream, the error bars are calculated from the error of the lag.
We give up the $\upsilon_{tan}$ from small lags of $\pm$2 s to avoid the spuriously large measurement errors. For the highly anisotropic fast stream, we display the velocity from both Ys-Hh and Mc-Hh.
Finally, the velocity obtained from spikes and cross-correlation is presented in Fig.\ref{fig:velocity_m1a09x}.\\
$\upsilon_{rad}$ of the CME front between 07:16-07:20 is
643$\pm$138 km s$^{-1}$.
When the FF become the most intense (07:20-07:30), the mean $\upsilon_{rad}$ is 480$\pm$40 km s$^{-1}$.
The scintillation pattern then is dominated by the declining velocity of CME material, 268$\pm$42 km s$^{-1}$ between 07:40-09:30, and 154$\pm$19 km s$^{-1}$ between 09:40-12:00. On the other hand, the fast streams display an acceleration progress, 725$\pm$25 $\sim$ 1106$\pm$50 km s$^{-1}$ obtained from Ys-Hh, 626$\pm$20 $\sim$ 848$\pm$31 km s$^{-1}$ from Mc-Hh.\\
$\upsilon_{tan}$ reverses its direction in our measurements, -514$\pm$119 km s$^{-1}$ equatorwards before 09:30, 84$\pm$9 km s$^{-1}$ polewards between 10:00-10:30, -290$\pm$70 km s$^{-1}$ equatorwards between 11:30-12:00, and 242$\pm$107 km s$^{-1}$ polewards between 12:30-13:00. \\

The oscillation of
 $\upsilon_{tan}$ exhibits large scale wavelike motions. The direction of $\upsilon_{tan}$ turns around between polewards and equatorwards, with the wave period of about 2 hours, and a propagation speed in the range of 60$\sim$600 km s$^{-1}$. The oscillation of the solar wind complies with the properties of the streamer waves, which is a decaying oscillation of the streamer after the CME passage~\citep{Chen2010,Decraemer2020}.
 \cite{Chen2010} and \cite{Decraemer2020} identify the streamer wave in the bright streamer belts from LASCO. The high density sensitivity and spatial resolution of our method enables to find the streamer waves near the north pole of the sun, a much dimmer area.\\
Comparing with the CME, where the lag and velocity can be obtained in all the baselines,
the propagation of the fast stream is super radial and highly anisotropic.
From Fig.\ref{fig:CCfreq}, the fast irregularities stretch out optimally along Hh-Ys, sub-optimally along Hh-Mc, non-significant propagation along Hh-Zc. It means the fast stream has a field-aligned anisotropy along the direction of Hh-Ys.
 In Fig. \ref{fig:radlatdis} and \ref{fig:impactpoint}, the Ys, Mc and Zc are different in tangential direction. Ys is $\sim$800 km equatorwards off Hh-Mc, and Zc is $\sim$1400 km polewards off Hh-Mc. Due to $\sim$2200 km deviation between Zc and Ys in tangential,
 the field-aligned anisotropy of the fast stream makes it undetectable along Zc-Hh.\\
The field-aligned anisotropy of the density fluctuations was also found in the angular broadening data from the Very Large Array (VLA) \citep{Armstrong1990,Grall1996}. \cite{Harmon2005} models the radio scattering and scintillation in the inner solar wind with the oblique
$\rm{Alfv\acute{e}n}$/ion cyclotron waves. It says the intensity scintillation (IPS) velocities show characteristics consistent
with the $\rm{Alfv\acute{e}n}$ wave dispersion relation. These characteristics include high field-aligned velocity spreads, low perpendicular velocity spreads.
It is possible that the field-aligned fast stream measured here is attributed to the effect from $\rm{Alfv\acute{e}n}$ waves, which propagates in the direction of the magnetic field. It is worthwhile to mention that the cross-correlation analysis presented here could be performed with Faraday rotation fluctuations to detect these magnetic field fluctuations as well as calculating the wave power contained within these fluctuations~\citep{Kooi2014}.\\
According to the data from Advanced Composition Explorer satellite (ACE)~\footnote{http://www.srl.caltech.edu/ACE/ASC/DATA/level3/icmetable2.htm}, the CME on Oct.9 is Earth-directed with the
Sun-Earth transporting velocity of 620 km s$^{-1}$. As an important supplement, we measure the velocity of CME at 2.62 Rs near the north pole of the sun. The
velocity of the CME front measured in the paper is consistent with ACE, and we also find the CME deflecting to ecliptic plane before 09:30. \\
\mml{The solar wind velocity measurement in this study has a rigorous requirement on the projected baselines directions formed by the radio telescopes. At least 4 telescopes are required to provide the special distribution. We are able to measure the radial velocity of the solar wind because the Hh related baselines cover a broad range in the projected radial distances, find the streamer wave because Zc-Mc, Mc-Ys and Zc-Ys cover a broad range in latitudinal distances, but a comparatively short range in radial distance, find the field-aligned fast stream because it happened to be highly anisotropic along Hh-Ys.
The IPS or FF power spectra analysis of spacecraft signals could also be used to infer the solar wind velocity in the corona~\citep{Imamura2014,Wexler2020}. These methods require only one telescope. However, the IPS focuses on the short-period waves around the Fresnel frequency. We can detect the field-aligned density irregularities caused by the propagation of $\rm{Alfv\acute{e}n}$ waves with longer period of 100-500 s. \cite{Wexler2019,Wexler2020} adopted the electron density model in their FF analysis, whereas, it's difficult for the model to depict the instantaneous variations of the electron density in the case of CME. We should combine the advantages of different methods to study the solar wind velocity in the future.}\\
\mml{This work is an in-beam observation of a satellite constellation, TIW and MEX. \cite{Kooi2022} referred that satellite constellations would provide multiple, closely-spaced ray paths to detect the solar corona. In this work, the CNR of MEX is strong enough to study the intensity fluctuations. We prepare to compare the multi-station cross-correlation method with the IPS method in the following work.}

\section{Conclusions} \label{sec:Conclusions}
With the reasonable distribution of VLBI radio telescopes, we firstly visually identify the drift of the scintillation patterns along the projected baselines in the radio band at the sky plane.
Combing the visual frequency spikes and the cross-correlation analysis, we have detected the variation of the solar wind velocity during a CME passage at a high temporal and spatial resolutions.
The oscillation of tangential velocity $\upsilon_{tan}$ confirms the detection of
streamer wave, which is usually found in bright streamer belts. At the tail end of the CME, we detect the field-aligned fast flow possibly relating to the $\rm{Alfv\acute{e}n}$ waves. The detailed physical interpretations of the oscillation and deceleration of the CME, as well as the field-aligned fast density fluctuations are still in research.\\
The FF observations of spacecraft by radio telescopes provide a unique source to characterize the nascent dynamic solar wind structure.
Besides the TIW and MEX, some other deep space spacecraft, e.g., the BepiColombo, the Mars Reconnaissance Orbiter, has a high quality beacon as well. We hope to further connect the radio and spacecraft to study the challenging inner solar wind in the future.\\

\acknowledgments
The authors would like to express our sincere gratitude to the Chinese and European space agency for providing the information of spacecraft.
We would like to express our sincere gratitude to the VLBI stations for them urgently
carrying out this and related observation(s). Their dedications and fraternities are admirable.
We would like to thank Prof. Fengchun Shu and Yong Huang for their important supports. We would like to express our sincere gratitude to the software correlator group for decoding the raw data, and many colleagues's friendly help in preparing the observations and transferring the data.
We would like to thank the anonymous reviewers for improving the quality of the paper.
We would like to thank the National Natural Science Foundation of China (grant numbers U1831137,41874205,U1931135,11703070) who funded the project.Thanks for scientific data by National Basic Science Data Center 'VLBI Radio Astronomy and Deep Space Exploration DataBase' (NO.NBSDC-DB-11). The work at Peking University is supported by NSFC (41874200, 42174194 and 42150105), and by National Key R\&D Program of China (2021YFA0718600).\\
\bibliography{IPS2022_2}

\appendix
\section{The projected baselines} \label{sec:projected}
\begin{figure}[htb]
\centering
\includegraphics[width=1.0\textwidth]{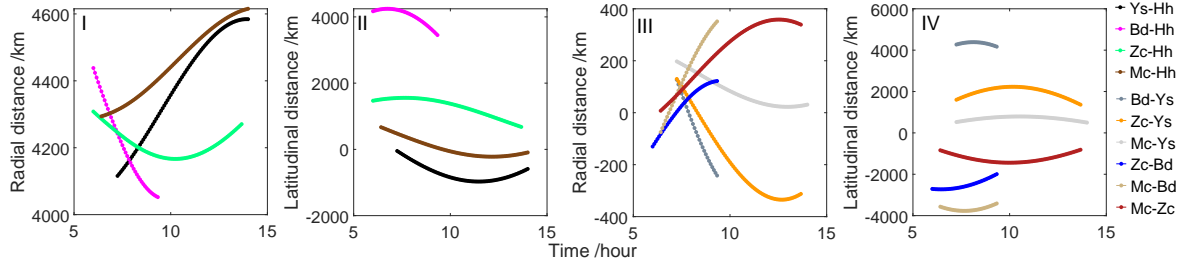}
\caption{The projected radial and latitudinal distance. I,II. Radial and latitudinal distance at Ys-Hh, Bd-Hh, Zc-Hh and Mc-Hh. III,IV. Radial and latitudinal distance at Bd-Ys, Zc-Ys, Mc-Ys, Zc-Bd, Mc-Bd and Mc-Zc.\label{fig:radlatdis}}
\end{figure}

\begin{figure}[htb]
\centering
\includegraphics[width=0.4\textwidth]{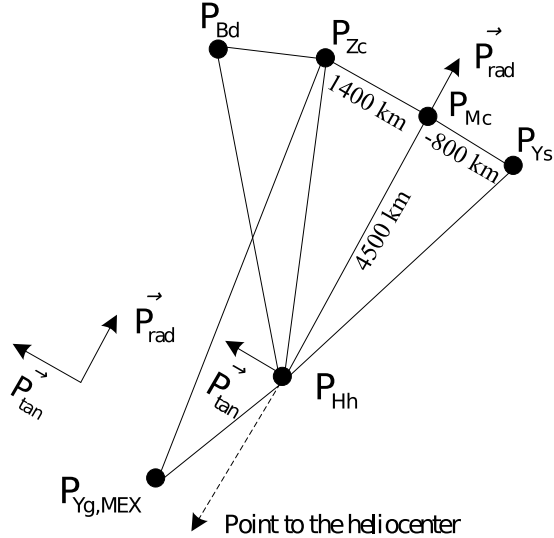}
\caption{The geometric diagrams of P-points in the sky plane on Oct.9. \mml{The distance are marked at UTC 11:00.} $P_{Yg,MEX}$ is the P-point closest to the sun from the LOS between MEX and Yg. Other P-Points are the closest point to the sun from TIW to the antennas.  \label{fig:impactpoint}}
\end{figure}

To study the propagation of solar wind, we have projected the baselines between the station-pairs onto the sky-plane in the heliographic coordinates, then calculated the components in the radial, latitudinal and longitudinal directions from the Sun. The point of closest approach of the line of sight(LOS) to the Sun is referred to as projected P-Point (i.e. impact parameter). The Carrington longitude of P-Points is 258$^{\rm{o}}$,
with the longitudinal components of all the projected baselines are usually less than 200 km, therefore we only focus on the radial and latitudinal (usually called tangential in IPS) distance difference between the station-pairs, see Fig.\,\ref{fig:radlatdis}. \mml{The reference radial direction is along the heliocenter and Hh. The tangential direction is pointing to the north pole of the sun.
 The projected baselines between Ys-Hh and Mc-Hh cover a broad range in radial distances, but a
 comparatively short range in latitudinal distance.}
 In addition, the baselines between Bd, Zc, Mc, Ys are sensitive in the latitudinal direction. The latitudinal components between Bd-Zc, Zc-Mc and Mc-Ys are about 2500, 1200 and 800 km, respectively. Meanwhile, the radial components between these station-pairs are
less than 400 km.
\mml{Fig.\,\ref{fig:impactpoint} are the geometric diagrams of projected baselines in the sky plane. At UTC 11:00, Hh-Mc is aligning the radial direction when the tangential distance between Hh and Mc is 0. We marked the distance scales between Hh, Mc, Ys and Zc at this moment.}
The Hh or Yg related baselines are sensitive to the radial direction. The radial solar wind will arrive $P_{Yg}$ first, then $P_{Hh}$, at last other P-Points.\\
\end{document}